\documentclass[sigconf, authorversion]{acmart}

\AtBeginDocument{%
  }

\copyrightyear{2025}
\acmYear{2025}
\setcopyright{cc}
\setcctype{by-nc-nd}
\acmConference[CHI '25]{CHI Conference on Human Factors in Computing Systems}{April 26-May 1, 2025}{Yokohama, Japan}
\acmBooktitle{CHI Conference on Human Factors in Computing Systems (CHI '25), April 26-May 1, 2025, Yokohama, Japan}\acmDOI{10.1145/3706598.3713827}
\acmISBN{979-8-4007-1394-1/25/04}
\acmDOI{10.1145/3706598.3713827}

\acmSubmissionID{740}

\usepackage{threeparttable}

\begin{document}

\title{Generative AI Uses and Risks for Knowledge Workers in a Science Organization}

\author{Kelly B. Wagman}
\orcid{0000-0002-9969-6188}
\affiliation{%
  \institution{University of Chicago}
  \city{Chicago, IL}
  \country{USA}}
\email{kbwagman@uchicago.edu}

\author{Matthew T. Dearing}
\orcid{0000-0002-1270-2857}
\affiliation{%
  \institution{Argonne National Lab}
  \city{Lemont, IL}
  \country{USA}}
\email{mdearing@anl.gov}

\author{Marshini Chetty}
\orcid{0000-0003-0804-6110} 
\affiliation{%
  \institution{University of Chicago}
  \city{Chicago, IL}
  \country{USA}}
\email{marshini@uchicago.edu}

\renewcommand{\shortauthors}{Kelly B. Wagman et al.}

\begin{abstract}
Generative AI could enhance scientific discovery by supporting knowledge workers in science organizations. However, the real-world applications and perceived concerns of generative AI use in these organizations are uncertain. In this paper, we report on a collaborative study with a US national laboratory with employees spanning Science and Operations about their use of generative AI tools. We surveyed 66 employees, interviewed a subset (N=22), and measured early adoption of an internal generative AI interface called Argo lab-wide. We have four findings: (1) Argo usage data shows small but increasing use by Science and Operations employees; Common current and envisioned use cases for generative AI in this context conceptually fall into either a (2) \textit{copilot} or (3) \textit{workflow agent} modality; and (4) Concerns include sensitive data security, academic publishing, and job impacts. Based on our findings, we make recommendations for generative AI use in science and other organizations.
\end{abstract}




\maketitle

\section{Introduction} 

Generative AI, as a new core technology, has demonstrated significant potential to enhance scientific discovery by supporting knowledge workers in science organizations and institutions (e.g., government and industry research labs, universities, think tanks) \cite{MSAI3Science_2023, 10Rules2024, Song_etal_2023}. However, the real-world applications and perceived risks of generative AI use in these organizations are uncertain. If generative AI could make science organizations with science- and operations-focused employees more efficient and speed up time to discovery on topics such as drug development and climate solutions, it would have important consequences for facilitating scientific gains to help society at large.

Prior literature has looked at generative AI as a tool for scientific research, including the development of science-specific large language models \cite{Song_etal_2023, Taylor_etal_2022} and driving complex scientific tasks \cite{Diao_microscope_2024, Lei_materials_science2024, Dearing_Prince_2023, Schick_2023}. This research, however, focuses on science research tasks and does not study the science workplace more broadly. Another area of literature looks at generative AI in the workplace, with a particular focus on professional knowledge workers \cite{Kaddour_etal_2023, Dwivedi_etal_2023, Sako_2024, Shi_Jain_Doh_Suzuki_Ramani_2024, Goth_2023, Woodruff_etal_2024, Butler_etal_2023}. Very little research, however, studies generative AI opportunities for scientists as knowledge workers \cite{Morris_2023}, and to the best of our knowledge prior research has not studied generative AI for both science- and operations-focused workers in a science organization. In addition, prior work has investigated generative AI risks and concerns (e.g., \cite{Dwivedi_etal_2023, Passi_Vorvoreanu_2022, Perry_Srivastava_Kumar_Boneh_2023, Morris_2023}), however, there has not been a focus on concerns that are specific to the novel context of a science organization.

In this paper, we add to the literature by reporting on an investigation of the practical, real-world applications and perceived risks of generative AI use (focusing predominately on large language models as opposed to image-generating models) across a multidisciplinary science and engineering research center with the primary mission to deliver scientific progress. To conduct our study, we collaborated with an Information Technology (IT) group at a US national laboratory.\footnote{US national labs are funded by the government but run by private corporations with which the government contracts.} The national lab, Argonne, we worked with employs several thousand people and is tasked with basic and applied science and engineering research. The lab includes Science divisions working to publish academic papers in research areas such as climate science, materials science, high performance computing, and more. Science teams also include software engineers, data scientists, and engineers who specialize in building and operating scientific experiments. In addition, the lab has multiple Operations divisions that employ people in areas such as IT, Human Resources (HR), Finance, Communications, Facilities (e.g., grounds crew and building maintenance), the Fire Department, and Security. We studied generative AI perceptions and uses for employees spanning Science and Operations roles and measured early adopter usage for a recently released internal generative AI interface called Argo based on a private instance\footnote{User data are not shared with any third-party external service. User queries and LLM responses are not stored through any system internal or external to the organization.} of OpenAI's GPT-3.5 Turbo large language model (LLM). We study all usage during the first several months Argo was available, and use the term \textit{early adopter} to highlight that we study these initial users of the system.

Two features of the national lab make it a useful case study for science organizations and non-science organizations interested in using generative AI. First, numerous knowledge work organizations have a similar dichotomy between knowledge specialists (e.g., scientists, lawyers, investors) and operations workers, and in this paper we consider that they might have different needs but similar levels of interest with respect to generative AI. Second, the national lab regularly deals with sensitive data, such as classified, national security, or pre-published scientific data, and thus must take privacy and security risks seriously, similar to organizations such as banks and other government institutions.

Our research questions are as follows:
\begin{enumerate}
    \item[\textbf{RQ1:}] How are Science and Operations workers at a national lab currently using, and how do they envision using, generative AI to support their work?
    \item[\textbf{RQ2:}] What risks (including privacy, security, and ethics) exist for using generative AI at a national lab?
\end{enumerate}
To answer our research questions, we conducted a survey (\(N=66\)) and in-depth semi-structured interviews (\(N=22\)) with Argonne employees, and also analyzed Argo usage data from the first eight months of deployment (which overlapped with the study time frame).

We have four main findings: (1) Following its initial launch, we found that Argo was being used by a growing number of early adopters (more often in Science) and that most survey respondents were familiar with, and experimenting using, generative AI. However, few had made it a consistent part of their work; We identified common generative AI use cases that conceptually binned into either a (2) \textit{copilot} or (3) \textit{workflow agent} modality. \textit{Copilot} refers to a system that works in conjunction with a user on tasks and provides responses in a conversational manner. \textit{Workflow agent} refers to an autonomous or semi-autonomous AI system that can perform complex tasks mostly on its own to support a user's work. Science and Operations participants reported similar copilot-style interaction scenarios: current uses centered around writing verifiable and structured code or text, while envisioned uses focused on extracting insights from large, unstructured text data. Both Science and Operations participants described how generative AI agents could automate some of their workflows, although these workflows differed between the two roles; and (4) In terms of generative AI risks, we found many participants were concerned about reliability/hallucinations, overreliance on generative AI, data privacy and security, the future of academic publishing and citation practices, and to what extent generative AI would impact hiring and jobs at the national lab.

Our paper's \textbf{novel contribution lies in studying the intersection of generative AI in scientific knowledge work and generative AI in a professional organization}. Specifically, we contribute: 
\begin{itemize}
    \item Novel usage data for the deployment of an organization-wide generative AI chatbot at a national lab (powered by a private instance of OpenAI's models).
    \item A collection of current and envisioned generative AI use cases for Science and Operations employees at a science organization binned into either copilot or workflow agent interaction modalities. 
    \item Novel perspectives from Science and Operations employees on generative AI risks for a science organization.
    \item Design recommendations for organizational use of generative AI copilots and workflow agents.
    \item Recommendations for future HCI research on generative AI applications in science, and other knowledge work, organizations.
\end{itemize}

\section{Background}

\subsection{Generative AI in Science}

In recent years, scientists have been exploring how deep learning and generative AI might be useful for advancing scientific research \cite{Sanger2024, MSAI3Science_2023, 10Rules2024}. Some have called for building science-specific large language models (LLMs) trained only on science literature \cite{Song_etal_2023, Taylor_etal_2022}. Another area of interest has been in the use of LLMs to generate science-focused programming code, including calculation kernels executed on HPC systems and parallel code snippets to train LLMs \cite{Nichols2024, Godoy2023, Dearing_Tao_Wu_Lan_Taylor_2024}. Generative AI is also being leveraged to create synthetic user research data for Human-Computer Interaction (HCI) research \cite{Hamalainen_Tavast_Kunnari_2023, Park_OBrien_Cai_Morris_Liang_Bernstein_2023} and drive automated process workflows for scientific tasks and experimental equipment \cite{Schick_2023, Lei_materials_science2024, Diao_microscope_2024, Dearing_Prince_2023}. 

Little research exists, however, on how generative AI is impacting the day-to-day aspects of scientific knowledge work. One exception is a study by Morris \cite{Morris_2023} who interviewed 20 scientists, at a variety of institutions, and identified generative AI opportunities as well as concerns, ranging from literature reviews and data analysis to applications in higher education given many participants' roles as university professors. Our paper differs from Morris' by focusing on scientists at an organization that has a science mission (as opposed to university professors and scientists working in the tech industry). Notably, we include the perspective of Operations workers, investigate organization-level trends, and specifically seek perspectives on privacy and security. Another closely related study did not exclusively focus on generative AI. In this work, Crosby et al. \cite{Crosby_etal_2023} use a human-centered methodology to inform the design of a suite of tools for ocean scientists that leverage ML to process image data. Finally, some work has looked at the ability for LLMs to summarize academic papers for scholars \cite{Glickman_Zhang_2024, Wang_Huang_Yan_Xie_He} or create data visualizations \cite{Maddigan_Susnjak_2023}. We expand on these studies by contributing a comprehensive look at how both Science and Operations staff in a science organization might use generative AI in their work and what concerns they have.

\subsection{Generative AI in Knowledge Work}

Researchers have also begun investigating the use of generative AI in knowledge work, a classification of labor involving the production of information-driven products and services as the key economic output \cite{Woodruff_etal_2024, Sako_2024}. Some professions considered knowledge work include data science, law, marketing, and finance \cite{Sako_2024, Woodruff_etal_2024}. Given generative AI's ability to process text-based information, there has been growing interest in how this technology will impact the jobs of professional knowledge workers \cite{Kaddour_etal_2023, Dwivedi_etal_2023, Sako_2024, Shi_Jain_Doh_Suzuki_Ramani_2024, Goth_2023, Lohr_2024, Loten_2023, Woodruff_etal_2024, Butler_etal_2023}. 

One thread of research looks to measure the productivity gains by knowledge workers who have access to generative AI tools \cite{Jaffe_etal_2024}. Multiple studies have found that generative AI tools increase productivity more for less skilled workers \cite{Brynjolfsson_Li_Raymond_2023, Noy_Zhang_2023}, including for workers in consulting \cite{DellAcqua_McFowland_2023} and programming \cite{Kazemitabaar_etal_2023, Peng_Kalliamvakou_Cihon_Demirer_2023, Coutinho_etal_2024}. Other research on generative AI applications in professional settings has largely focused on creative knowledge work in journalism and science writing \cite{Petridis_etal_2023, Kim_Suh_Chilton_Xia_2023}, user experience (UX) and industrial design \cite{Uusitalo_etal_2024}, marketing and public relations \cite{Sun_Jang_Ma_Wang_2024}, and software engineering \cite{Russo_2024, Coutinho_etal_2024}. Another research direction has investigated creative knowledge work tasks (rather than professions) that include writing \cite{Gero_Calderwood_Li_Chilton_2022, Lee_etal_2024} or music composition \cite{Suh_Youngblom_Terry_Cai_2021}. We contribute to the literature on generative AI in knowledge work by focusing on science as a sub-domain that remains understudied.

In addition to a lack of research on scientific knowledge work, there is also minimal literature in the HCI and Computer Supported Cooperative Work and Social Computing (CSCW) communities studying generative AI at the organizational level. This oversight is significant because there can be networked impacts to organizations (both positive and negative) that might not be felt by individual users. For example, Cortiñas-Lorenzo et al. \cite{Cortinas-Lorenzo_Lindley_Larsen-Ledet_Mitra_2024} think critically about how enterprise knowledge systems accessed via AI have the ability to shape and distort how workers view themselves and others, which in turn changes worker behavior. One scenario they describe is a system that allows workers to search for ``experts'' on a topic in their organization. By labeling some employees experts and others not the system might inadvertently inform who gets promoted or increase workload on certain staff, making it imperative organizations make these systems transparent to workers. We contribute an in-depth case study of a single organization's use of generative AI for knowledge work, considering the interplay between roles (Science and Operations) within the organization as well as risks at the organizational level.

\subsection{Generative AI Concerns in the Workplace}

Alongside the search for generative AI applications in the workplace, researchers have uncovered risks, harms, and concerns relevant for knowledge workers. One of the most prevalent is the fact that generative AI systems ``hallucinate'' false information with suggestive confidence that it is true \cite{Kobiella_etal_2024, Dwivedi_etal_2023}. These hallucinations, as well as a lack of citations to original sources, can be particularly concerning for scientists \cite{Morris_2023}. There is also concern about overreliance, or users placing too much trust that a system is working and not checking outputs thoroughly enough \cite{Passi_Vorvoreanu_2022, Dwivedi_etal_2023}. Overreliance is a concern that predates the development of generative AI extending to early automation systems, such as automated landing systems for airplanes \cite{Sellen_Horvitz_2024}. Thus, researchers are considering how to design for the appropriate level of trust users \textit{should} put in a system \cite{Wang_Cheng_Ford_Zimmermann_2024}. 

Privacy and security as well as copyright and plagiarism are also work-related generative AI concerns. Some research has found that programmers who used a generative AI coding assistant were more likely to write insecure code \cite{Perry_Srivastava_Kumar_Boneh_2023, Pearce_etal_2022}. Additionally, significant controversy surrounds the use of writing \cite{Reisner_2023}, music \cite{Carras_2024}, and visual art \cite{Heikkila_2023, Shan_Ding_Passananti_Wu_Zheng_Zhao_2024} to train generative AI, due to the lack of compensation for artists and authors as well as the ability for the models to regurgitate some training data verbatim triggering copyright laws.

Another important issue drawing commentary from both researchers and the public is the extent to which generative AI will impact workers and industries, particularly risks to knowledge workers' jobs \cite{Woodruff_etal_2024, Lohr_2024, Goth_2023}. For instance, Woodruff et al. \cite{Woodruff_etal_2024} interviewed knowledge workers in several industries (their sample did not include scientists) about how they thought generative AI might change their field. They found that while participants did not foresee major changes to their industries, there was concern for the rise of deskilling, dehumanization, disconnection, and disinformation. Other research found workers can feel inferior or unaccomplished if generative AI models can largely do their tasks for them \cite{Kobiella_etal_2024}. While the spectre of fully autonomous systems has loomed for many industries going back to the Industrial Revolution, there have always remained some aspects of a job that machines cannot do \cite{Gray_Suri_2019}. In this paper, we expand on some of these risks and concerns while highlighting how they surface in unique ways both in a science context as well in an organizational context.

\section{Methods}

\subsection{Research Context}
In January 2024, the IT group at Argonne National Lab broadly deployed a generative AI chatbot named Argo, powered by a private instance of OpenAI's GPT-3.5 Turbo,\footnote{Argo was upgraded with additional features and access to GPT-4 Turbo after data collection for this study was complete.} for use by all members of the Argonne National Lab community. Argo was designed for exclusive internal lab use so it does not save query and LLM response data and it does not share such information with OpenAI or other third-party services. Employees could use Argo as they would a service like ChatGPT: it included a browser-based interface that a user could type a prompt into and get a reply. Argo could also be accessed by employees via an API. Argo was intended to be used only for work purposes and required a lab login and VPN to access if not on-site at the lab. Beyond providing a secure generative AI assistant for employees, Argo was not released with a specific purpose. Employees could find out about Argo through official announcements, emails, or events targeted at raising awareness.

Argonne National Lab is organized into Science and Operations divisions. The goal of Science divisions is to produce scientific research, while the goal of the Operations divisions is to keep the organization running smoothly. Science workers may be operating instruments, running experiments, managing data analysis, writing grants, and publishing papers. Operations workers may be producing science communication, working in administrative roles, ensuring lab safety (e.g. radiation exposure, cybersecurity), and building software and hardware infrastructure. Importantly, while the lab is divided into distinct divisions there is overlap in tasks between them. For example, both divisions employ technical workers such as software engineers. In addition, both divisions require some workers with a scientific background. In this paper, we distinguish between the two divisions to tease out if and when the need for generative AI differs between Science and Operations roles since this is a common division of labor in knowledge work organizations. 

We conducted a survey (April--June 2024) and interviews (April--July 2024) during the first eight months of Argo's deployment to purposefully capture generative AI perceptions and use levels during this early stage. Our research was approved by the university Institutional Review Board (IRB). Since it is against the law to monetarily compensate federal employees in the course of their work, we did not provide gift cards to participants, but made it clear the research would be used by the lab to improve their systems.

\begin{table*}%
\footnotesize
\begin{threeparttable}
\caption{Survey Participant Demographics}
\label{table:survey_participants}
\Description{Table includes survey participant demographics data. Participants are split fairly evenly between Science and Operations roles and skew towards white men, which reflects the demographics of the lab.}
\centering
\begin{tabular}{l|l|l|l} 
\hline
\textbf{Demographic} & \textbf{Response Option} & \textbf{Number of Participants (N=66)} & \textbf{Percentage}\\
\hline
\textbf{Division} & Science & 32 & 48\% \\ 
  & Operations & 31 & 47\% \\
  & Prefer not to answer & 3 & 5\% \\
\hline
 \textbf{Role} & Scientist & 15 & 23\% \\ 
  & Software Engineer, Data Scientist, IT  & 15 & 23\% \\
  & Operations Manager & 9 & 14\% \\
  & Engineer (hardware, facilities, etc.)  & 7 & 10\% \\
  & Cybersecurity, Safety  & 6 & 9\% \\
  & Scientist Manager & 5 & 7\% \\
  & Administrative, Communications  & 4 & 6\% \\
  & Prefer not to answer & 5 & 8\% \\
\hline
 \textbf{Years at lab} & <1 & 3 & 4\% \\
 & 1-4 & 17 & 26\% \\
 & 5-9 & 13 & 19\% \\
 & 10-14 & 9 & 14\% \\
 & 15-19 & 9 & 14\% \\
 & 20+ & 9 & 14\% \\
 & Prefer not to answer & 6 & 9\% \\
\hline
 \textbf{Age} & Under 24 & 1 & 1.5\% \\
 & 25-34 & 7 & 11\% \\
 & 35-44 & 20 & 30\% \\
 & 45-54 & 14 & 21\% \\
 & 55-64 & 13 & 20\% \\
 & 65-74 & 1 & 1.5\% \\
 & 75+ & 0 & 0\% \\
 & Prefer not to answer & 10 & 15\% \\
\hline
\textbf{Gender} & Female & 12 & 18\% \\
 & Male & 44 & 67\%\\
 & Non-binary & 0 & 0\%\\
 & Prefer not to answer & 10 & 15\% \\
\hline
\textbf{Race/Ethnicity} & American Indian or Alaska Native & 1 & 1.5\% \\
 & Asian & 5 & 7\% \\
 & Black or African American & 1 & 1.5\% \\
 & Hispanic, Latino, or Spanish Origin & 0 & 0\% \\
 & Middle Eastern or North African & 0 & 0\% \\
 & Native Hawaiian or Pacific Islander & 0 & 0\% \\
 & White & 46 & 70\% \\
 & Prefer not to answer & 13 & 20\% \\
\hline
\textbf{Education} & Less than high school degree & 0 & 0\% \\
& High school degree & 0 & 0\% \\ 
& Associate’s / Some college degree & 6 & 9\% \\
& Bachelor’s degree & 13 & 20\% \\
& Master’s degree & 20 & 30\% \\
& Doctoral degree & 21 & 32\% \\
& Prefer not to answer & 6 & 9\% \\
\hline
\end{tabular}
\end{threeparttable}
\end{table*}

\begin{table*}%
\footnotesize
\begin{threeparttable}
\caption{Interview Participant Demographics}
\Description{Table includes interview participant demographics. Participants are split fairly evenly between Science and Operations roles and skew towards white men, which reflects the demographics of the lab.}
\label{table:interview_participants}
\renewcommand{\arraystretch}{1.2} %
\centering
\begin{tabular}{l|l|l|l|l|l|l|l} 
\hline
 \textbf{ID} & \textbf{Division} & \textbf{Role} & \textbf{Years at lab} & \textbf{Age} & \textbf{Gender} & \textbf{Race/Ethnicity} & \textbf{Education}\\
 \hline 
P1 & Operations & Cybersecurity, Safety & 5-9 & 35-44 & M & White & Master’s degree\\
P2 & Operations & Engineer & 20+ & 55-64 & F & White & Master’s degree\\
P3 & Operations & IT & 10-14 & 35-44 & M & White & Bachelor’s degree\\
P4 & Operations & Cybersecurity, Safety & 1-4 & 35-44 & M & White & Bachelor’s degree\\
P5 & Operations & Engineer (Facilities) & 1-4 & 35-44 & M & White & Bachelor’s degree\\
P6 & Operations & IT & 1-4 & 55-64 & M & White & Associate’s / Some college degree\\
P7 & Operations & Cybersecurity, Safety & 5-9 & 25-34 & M & White & Associate’s / Some college degree\\
P8 & Operations & Operations Manager & 15-19 & 35-44 & F & White & Master’s degree\\
P9 & Operations & Cybersecurity, Safety & 15-19 & 35-44 & M & White & Master’s degree\\
P10 & Operations & Cybersecurity, Safety & 1-4 & <24 & F & White & Master’s degree\\
P11 & Science & Scientist Manager & 1-4 & 35-44 & M & White & Doctoral degree\\
P12 & Science & Scientist & 1-4 & 25-34 & F & Asian & Doctoral degree\\
P13 & Science & Scientist & <1 & 25-34 & M & White & Doctoral degree\\
P14 & Science & Data scientist & N/A & N/A & N/A & N/A & N/A\\
P15 & Science & Scientist Manager & 15-19 & N/A & M & N/A & Doctoral degree\\
P16 & Science & Scientist & 10-14 & 35-44 & M & White & Doctoral degree\\
P17 & Science & Scientist Manager & 10-14 & 55-64 & M & White & Doctoral degree\\
P18 & Operations & Operations Manager & 1-4 & 35-44 & F & White & Master’s degree\\
P19 & Science & Scientist & 5-9 & 35-44 & M & White & Doctoral degree\\
P20 & Science & Software Engineer & 15-19 & 35-44 & M & White & Bachelor’s degree\\
P21 & Science & Scientist Manager & 20+ & 45-54 & M & White & Doctoral degree\\
P22 & Science & Scientist & 20+ & N/A & M & N/A & Doctoral degree\\
\hline
\end{tabular}
\end{threeparttable}
\end{table*}

\begin{table*}%
\small
\caption{Themes* and Unique Survey Respondent Counts}
\Description{Table includes thematic codes as well as the number of unique survey respondents in each category. The three top-level codes are ``Current/envisioned uses for LLMs,'' ``Issues or ethics concerns at work,'' and ``Privacy/security concerns at work.''}
\label{table:codebook_survey}
\centering
\begin{tabular}{p{0.3\linewidth} | p{0.55\linewidth} | p{0.1\linewidth}} 
\hline
 Code & Code Description & Survey \newline Respondents (\%**) \\
\hline
\textbf{Current/envisioned uses for LLMs} &  \\
 Writing structured text/code & Use cases related to writing structured text/code & 13 (20\%) \\
 Query unstructured data & Use cases related to querying unstructured data sets & 13 (20\%) \\
 Workflow automation & Use cases related to workflow automation & 18 (27\%)\\
 Other & Other use cases & \hspace{.07cm} 5 (7\%) \\
\hline
\textbf{Issues or ethics concerns at work} & \\
Reliability & Concerns about whether LLMs are reliable, trustworthy, etc. & 29 (44\%) \\
Overreliance & Concerns about people relying too much on LLMs & 14 (21\%) \\
Academic publishing & Concerns about academic and science communication publishing & 13 (20\%) \\
AI taking jobs & Concerns about the impact of LLMs on human jobs & \hspace{.07cm} 3 (5\%) \\
Social concerns & Broader social concerns that participants do not tie directly to their work & \hspace{.07cm} 9 (14\%) \\
No ethics concerns at work & Participant did not have ethics concerns about LLMs at work & 22 (33\%) \\
\hline
\textbf{Privacy/security concerns at work} &  \\
Concerned & All concerns related to privacy and security & 28 (42\%) \\
Not concerned if using organization tools & Participant not concerned about privacy/security as long as they were using LLMs officially designated secure by their organization & 12 (18\%) \\
Concerned but prefer commercial LLMs & Participant felt organization-designated LLMs were safer but preferred the features available in commercial LLMs & \hspace{.07cm} 3 (5\%)\\
No privacy/security concerns at work & Participant did not have privacy/security concerns about LLMs at work & 22 (33\%) \\
\hline
\end{tabular}
\begin{tablenotes}
\item *Bolded codes correspond to the interview codebook codes that we analyzed for this paper. Sub-codes correspond to themes used to code the survey short responses.
\item **Because not every survey response included each theme, and some included multiple themes, percentages do not sum to 100\%.
\end{tablenotes}

\end{table*}

\subsection{Data Collection}

\subsubsection{Survey}

We first wanted to broadly capture national lab employee perceptions of generative AI uses and concerns in a survey. The survey, created in Qualtrics, included questions such as \textit{How familiar are you with large language models (LLMs)?}, \textit{How often do you use LLMs as part of your work?}, and \textit{Please describe ethical concerns you have about using LLMs in your work, if any}. We also asked \textit{How often do you use LLMs for the following work tasks?} and provided a list of 15 tasks such as \textit{writing code}, \textit{feedback on experimental design}, and \textit{editing human-written text}. This list of tasks was drawn from a study of 20 scientists about their use of generative AI \cite{Morris_2023}. We also asked demographics questions. At the end of the survey, we gave participants the option to share their email if they would be willing to sign up for an interview. See the full survey in Appendix \ref{appendix:survey}.\footnote{The full Appendix can be found in the supplementary materials.} We pilot-tested the survey with a small set of HCI experts in Computer Science and refined wording and ordering of the questions. After finalizing the changes, we deployed the survey to Argonne National Lab employees via mailing lists and at presentations given by our collaborators at the lab. 

We received a total of 80 survey responses. We filtered out surveys that were incomplete, with the exception of three that only lacked demographic information (given that a valid option was ``Prefer not to answer'' we selected this option for these responses) or short answer responses, but had all the multiple choice questions completed. We also removed one response that was disingenuous based on the answers. After filtering, we were left with 66 responses to analyze.

\subsubsection{Interviews}
As survey responses were completed, we contacted respondents who had provided email addresses for follow-up interviews. In total, we contacted 40 respondents, 22 of whom agreed to participate in semi-structured interviews. Each interview lasted 30 minutes and was conducted over Zoom and recorded. The interview protocol was designed to elicit more depth on generative AI applications, risks, and concerns than the survey. We asked participants to recall current or envisioned scenarios for generative AI, and also prompted them based on their survey responses. We then went in-depth on each scenario, asking questions such as \textit{To what extent have LLMs been helpful for this task?} and \textit{When using an LLM to do this task doesn't work, what goes wrong?} In addition, we dug into participant views on privacy, security, and ethics concerns. See our full interview protocol in Appendix \ref{appendix:interview_protocol}.

\subsubsection{Argo Usage Statistics} 
Between January and August 2024, we collected high-level metadata during the regular use of Argo including authenticated user name, time of use, LLM selected and related configuration options, and the size of user queries and responses as measured by token counts.\footnote{A \textit{token} is unit of data processed by an LLM, such as a word, a part of a word, or a collection of symbols. A count of the number of tokens representing a block of text is a simple and standard measure of the size of that text, without exposing its content.} The text-based content of these queries and responses are not automatically stored by any external or internal database system. Only an Argo user  may optionally save their LLM-based conversations to their local machines before disconnecting from their session with the service through a copy-to-clipboard mechanism or a transcript download feature. We roll-up the tracked user name to their organizational division, which is then categorized as either being within the Science or Operations workforce group. 

\subsection{Participants}
Table \ref{table:survey_participants} shows the demographic information for the survey participants. We received approximately equal numbers of Science and Operations employee responses. Most respondents were scientists or engineers.  Respondents had been at the lab for a range of years, including both new and highly seasoned employees. We also represent a range of ages, primarily between 25 and 64 years old. The gender and race/ethnicity breakdown is representative of the lab population. Most (but not all) respondents had at least a Bachelor's degree, and many had advanced degrees.

Table \ref{table:interview_participants} shows demographic information for interview participants. All interview participants also provided a survey response, and the demographic breakdown is similar. It is skewed toward white men, although this is reflective of the overall national lab population. Throughout the paper, we denote interview participants IDs with ``P'' and survey participant IDs with ``S.'' 

\subsection{Data Analysis}

\subsubsection{Interviews}
We had the interviews transcribed using Rev.com under a non-disclosure agreement. We qualitatively coded the interview transcripts and then analyzed them using thematic analysis \cite{saldana_coding_2013}. First, one researcher read through the transcripts and created an initial codebook based on the interview questions. The team then discussed and refined the codebook and finalized the codes (the codes we analyzed for the final paper are bolded in Table \ref{table:codebook_survey}, see Table \ref{table:codebook_interview} in Appendix \ref{appendix:codebook_interview} for the full initial codebook). One researcher subsequently coded all the transcripts using MAXQDA. Then, the same member of the research team used axial coding to assign sub-codes through an iterative process to the following parent codes: \textit{Current/envisioned use cases for LLMs}, \textit{Privacy/security concerns at work}, and \textit{Issues or ethics concerns at work}. Another member of the research team then did a second round of coding, reviewing all sub-codes labeled in the first round. All points of disagreement were discussed and resolved between the two coders. The team also met regularly after a set of transcripts were coded to ensure the coding process aligned with the goals of the study. After coding, two members of the research team extracted themes and discussed these with the full team (see themes in Table \ref{table:codebook_survey} as well as Table \ref{table:codebook_interview} in Appendix \ref{appendix:codebook_interview}) and further refined those that were most related to the research questions. 

\subsubsection{Survey}
We calculated descriptive statistics for the survey based on the multiple choice responses using Python. We qualitatively coded the short answer responses in MAXQDA. We began the coding process after extracting themes from the interview data. Thus we used the same codebook for the survey, as shown in Table \ref{table:codebook_survey}, along with the unique number of survey responses for each code.

\section{Findings}

Our findings represent the perspectives of generative AI early adopters at the national lab. In Section \ref{findings:usage}, we describe initial generative AI usage at the lab, showing that less than 10\% of all lab employees used Argo each month in the study period but that there is an upward trend in use. This is a lower bound for generative AI usage more broadly because in our surveys, more employees reported trying other commercial LLMs such as ChatGPT. 

Drawing from both the survey short responses and the interviews, we identified common \textit{current} and \textit{envisioned} use cases for generative AI and conceptually split these findings into two categories: \textit{copilot} (Section \ref{findings:copilot}) and \textit{workflow agent} (Section \ref{findings:agent}). We review similarities and differences between Science and Operations workers with respect to the kinds work they want to accomplish in each category.  In Section \ref{findings:risks}, we highlight the most common risks and concerns for using generative AI in the national lab that surfaced in both the survey short answers and interviews.

\subsection{Generative AI Usage and Familiarity}
\label{findings:usage}    

\begin{figure*}%
    \centering
    \includegraphics[width=.95\textwidth]{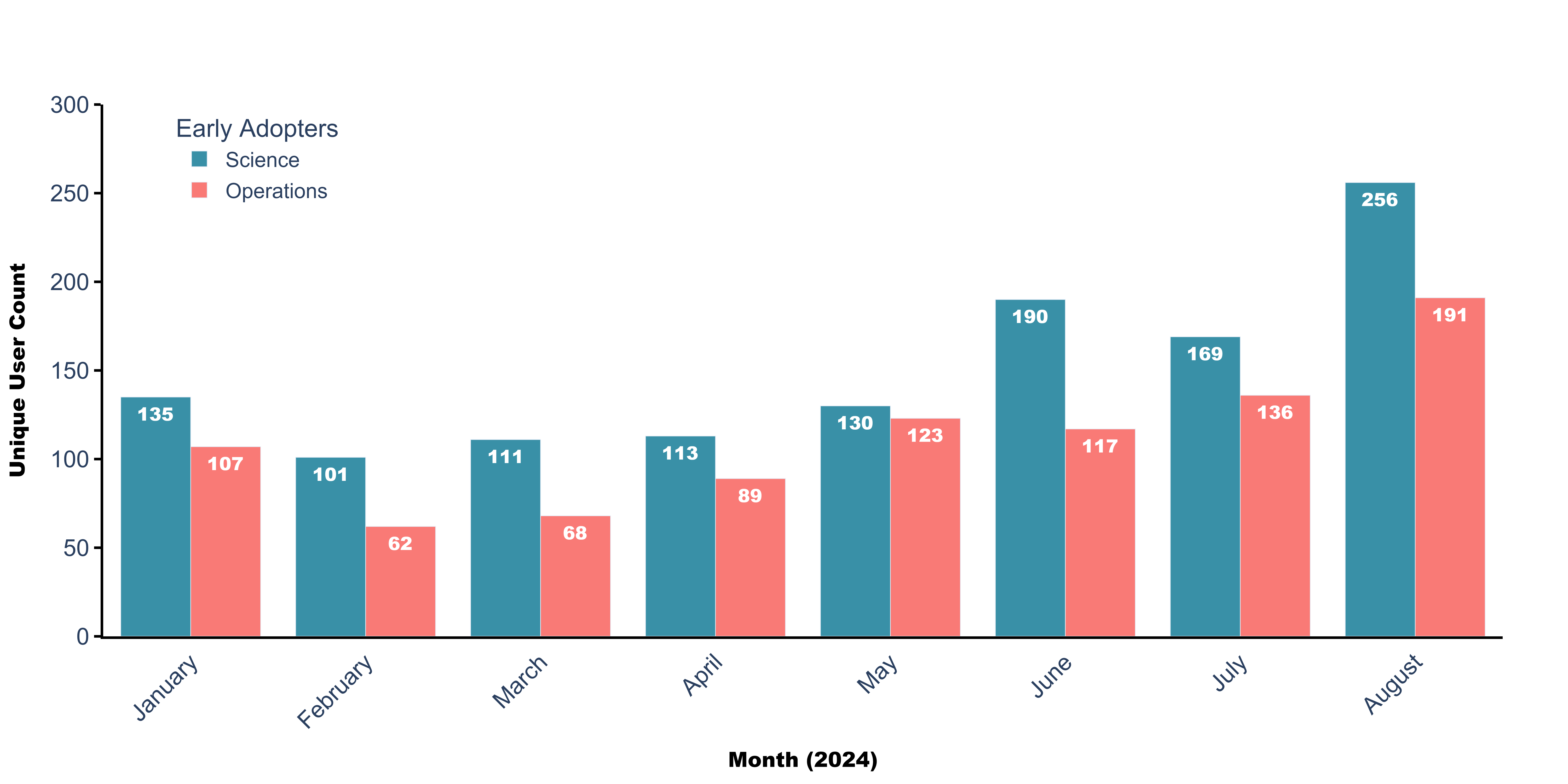}
    \caption{Argo usage metrics by Operations and Science unique users for each month since initial deployment. Monthly usage is less than 10\% of all lab employees. \textit{Note: this plot does not capture employee use of external LLMs and is based on auto-collected telemetry data.}}
    \Description{Figure represents Argo usage metrics by Operations and Science unique users each month, from January to August 2024. For each month, there are more Science than Operations users, although both roles show an upward trend over time.}
    \label{fig:ops_sci_use_trends}
\end{figure*} 

We found that early adopters are familiar with generative AI and using it experimentally in their work from our survey data and Argo usage data. Figure \ref{fig:ops_sci_use_trends} shows an upward trend in usage\footnote{The usage data is for all users, which may include survey respondents but does not correspond directly to them.} during the early months of launching Argo. While the use period seen in Figure \ref{fig:ops_sci_use_trends} is brief for designating trends, we observed a general increase in generative AI use across Science and Operations during the study period. Specifically, unique users of both groups, excluding the initial launch month, increased an average of 19.2\%  each month, broadly ranging from approximately no change to as high as nearly 47\%. Across this same period, monthly unique users in Science and Operations increased 158\% and 200\%, respectively (or 174.2\% over all users between the second and final months of the reported statistics). While there were 26\% more unique monthly Science users on average compared to Operations users, we observe an \textit{average monthly increase} in the latter group at 21.2\% compared to 19.3\% for Science users. Overall, we see a suggestive early trend that the availability of a secure internal generative AI solution is of significant interest to Science and Operations users at a national lab, and this data provides a useful baseline for future studies. 

\begin{figure*}%
    \centering
    \includegraphics[width=0.9\textwidth]{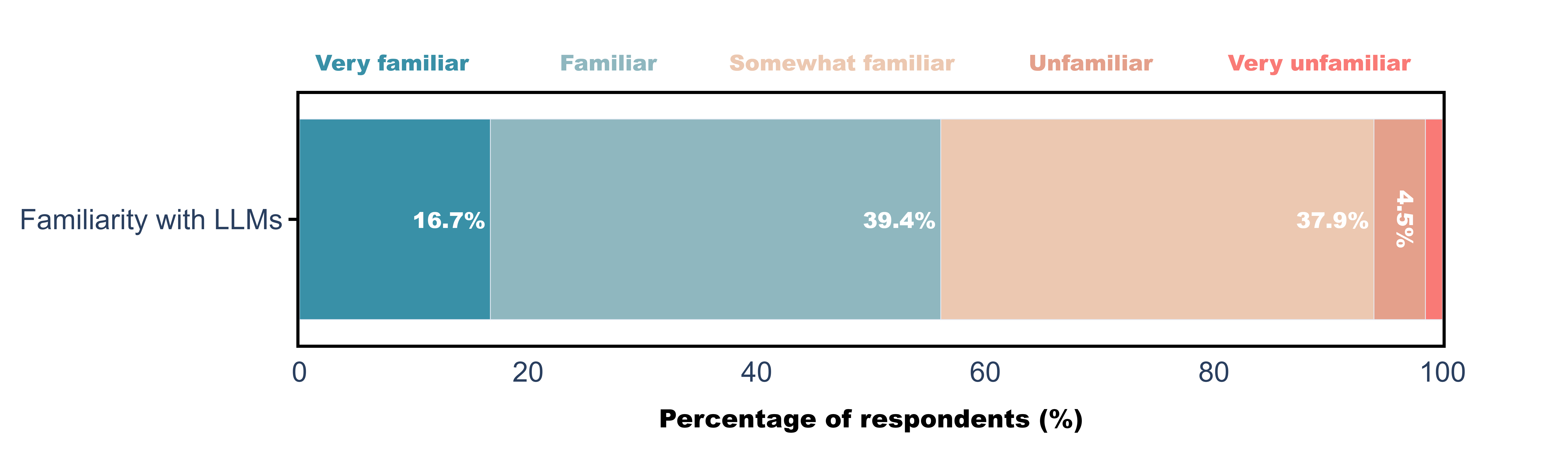}
    \caption{Responses to the survey question: \textit{How familiar are you with large language models (LLMs) such as ChatGPT, Argo, etc.?}}
    \label{fig:familiarity}
    \Description{Figure represents distribution of answers to the survey question: How familiar are you with large language models (LLMs) such as ChatGPT, BOT_NAME, etc.? Most participants had some level of familiarity with generative AI tools, but were not ``very familiar.''}
\end{figure*}

\begin{figure*}%
    \centering
    \includegraphics[width=0.9\textwidth]{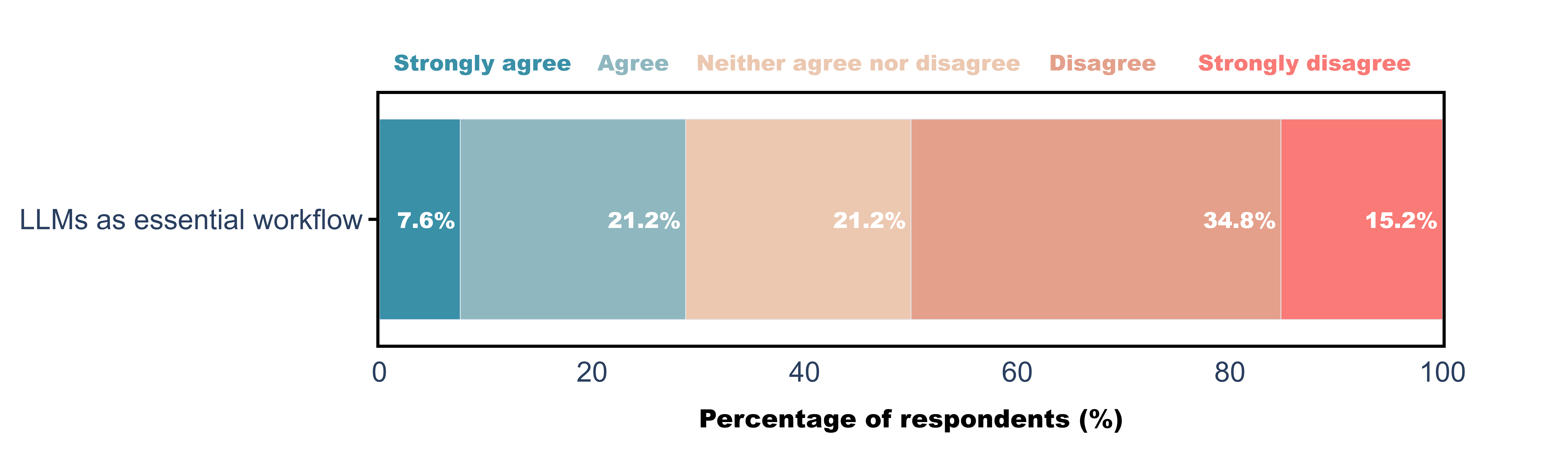}
    \caption{Responses to the survey question: \textit{To what extent would you agree/disagree with the following statement: LLMs have become an essential part of my workflow.}}
    \Description{Figure represents distribution of answers to the survey question: To what extent would you agree/disagree with the following statement: LLMs have become an essential part of my workflow. About a third of survey respondents said they agreed or strongly agreed that LLMs had become essential to their workflows.}
    \label{fig:workflows}
\end{figure*}

\begin{table*}%
\small
\caption{Responses to the survey question: \textit{Which LLMs have you used before?}}
\Description{Table includes a list of large language models and number of respondents who have used each. The most common large language model is ChatGPT, followed by Argo.}
\label{table:llms_used}
\centering
\begin{tabular}{l | l} 
\hline
 LLM & Respondents (\%) \\
\hline
ChatGPT (OpenAI) & 54 (82\%) \\
Argo & 44 (67\%) \\
Gemini (Google) & 17 (26\%) \\
LLaMA (Meta) & 11 (17\%) \\
Claude (Anthropic) & 6 (9\%) \\
Other & 7 (11\%) \\
None & 7 (11\%) \\
\hline
\end{tabular}
\end{table*}

\begin{figure*}%
    \centering
    \includegraphics[width=0.99\textwidth]{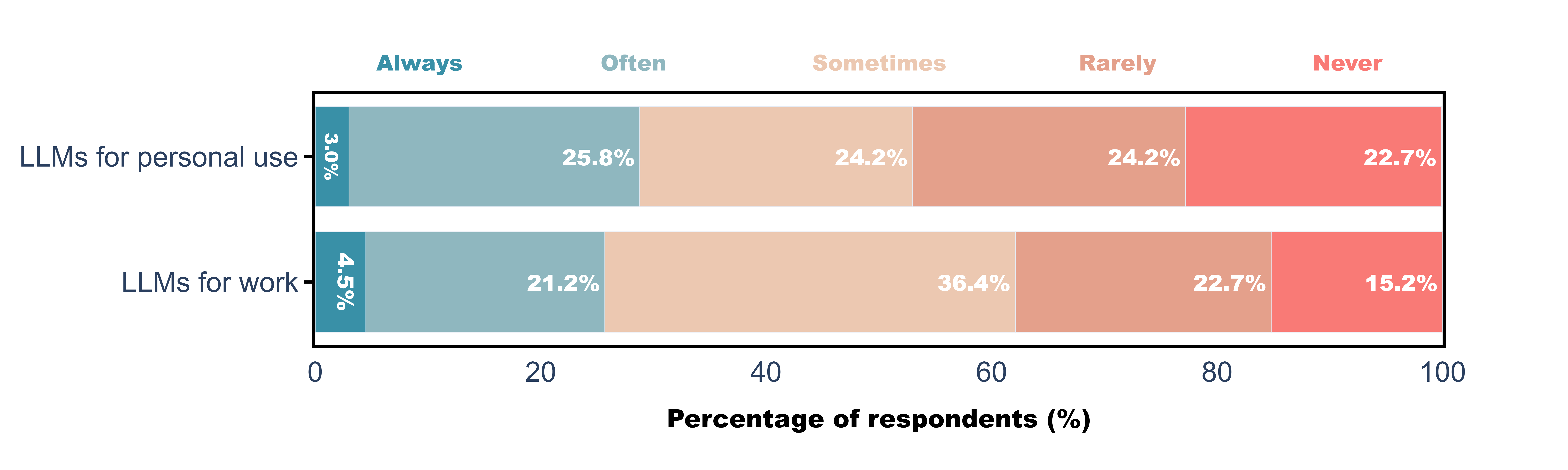}
    \Description{Figure represents distribution of answers to the survey questions: How often do you use LLMs as part of your work? and How often do you use LLMs for personal use? There was not a large difference between how much respondents used LLMs for work and personal tasks.}
    \caption{Responses to the survey questions: \textit{How often do you use LLMs as part of your work?} and \textit{How often do you use LLMs for personal use?}}
    \label{fig:personal_v_work_tasks}
\end{figure*}

\begin{figure*}%
    \centering
    \includegraphics[width=0.99\textwidth]{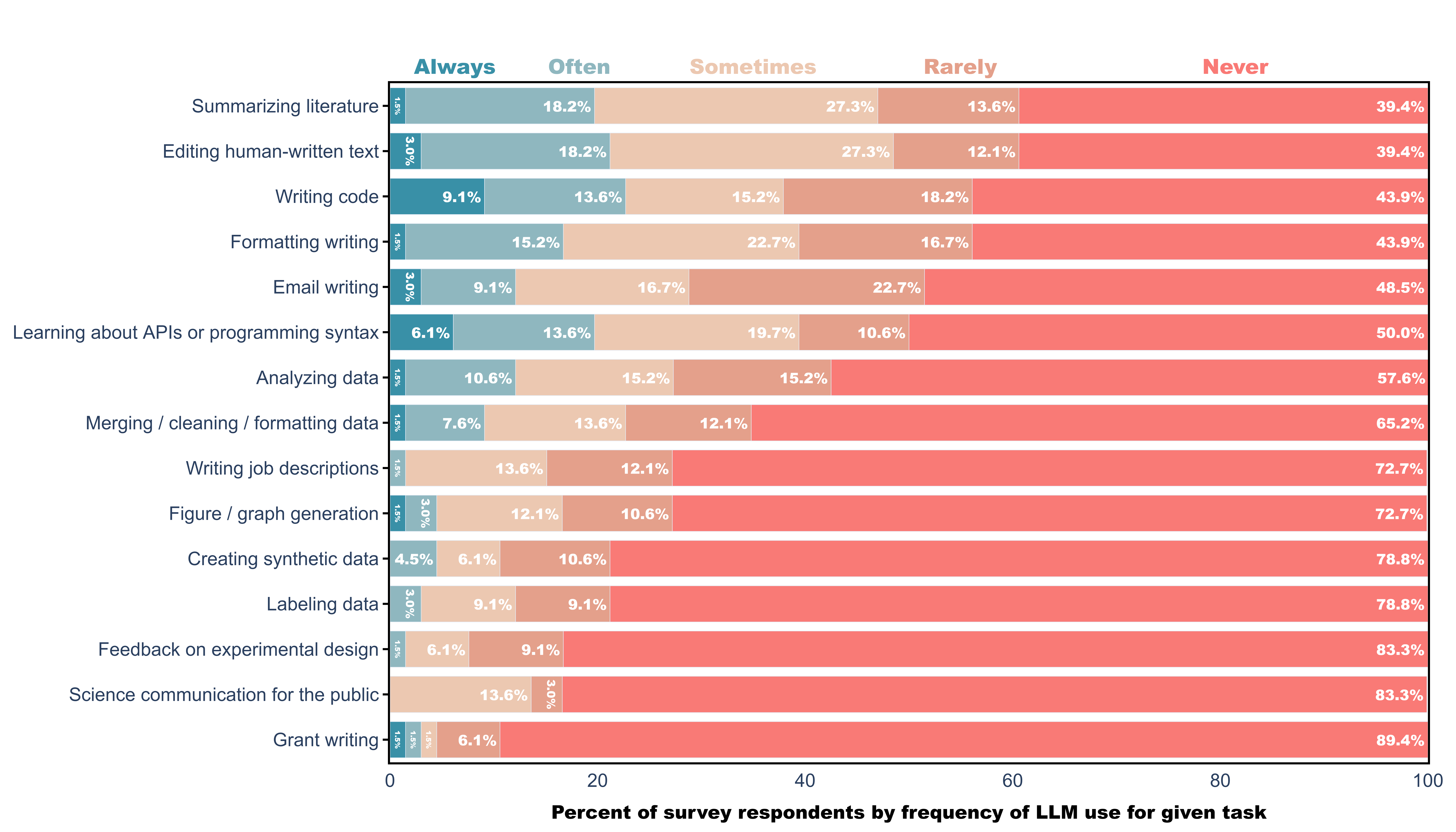}
    \caption{Responses to the survey question: \textit{How often do you use LLMs for the following Argonne National Lab-related work tasks?}}
    \Description{Figure represents distribution of answers to the survey question: How often do you use LLMs for the following Argonne National Lab-related work tasks?, with answers for each of 15 specific tasks. Respondents appear to be in an experimentation phase, and over half had at least tried using LLMs for summarizing literature, editing human-written text, writing code, and formatting writing.}
    \label{fig:frequencytasks}
\end{figure*} 

Most of our survey respondents, and all our interview participants, had some degree of familiarity or experimentation with generative AI. Based on the survey short responses and interviews, participants also had a sophisticated understanding of how generative AI models are trained. In our surveys, we asked employees about their familiarity with, and use of, LLMs more broadly. Figure \ref{fig:familiarity} shows that 94\% of participants had some level of familiarity with generative AI tools, although only 16.7\% felt they were \textit{very familiar}. Similarly, despite high levels of familiarity, only 28.8\% of survey respondents said they \textit{agreed} or \textit{strongly agreed} that LLMs had become essential to their workflows (Figure \ref{fig:workflows}). The majority of  survey respondents had primarily tried ChatGPT (82\%) and Argo (67\%) as shown in Table \ref{table:llms_used}. Our survey respondents also did not report a large difference between how much they used LLMs for work and personal tasks as shown in Figure \ref{fig:personal_v_work_tasks}. 

Figure \ref{fig:frequencytasks} shows how often survey respondents used LLMs for a series of tasks in the context of their work. Nearly 60\% of respondents had at least tried using LLMs for \textit{summarizing literature}, \textit{editing human-written text}, \textit{writing code}, and \textit{formatting writing}. While the fewest total respondents (11\%) had tried using LLMs for \textit{grant writing}, we note that one person responded \textit{always}, suggesting that it might be that respondents want to use LLMs for these tasks but have not figured out how to given their relative complexity. Examples of personal generative AI use that came up un-prompted in the interviews included scripting narratives as part of the game Dungeons and Dragons (P5, P13, P19) and creating furniture arrangements for a floor plan (P12). This data suggests early adopters are in an experimentation phase. 

\bgroup
\raggedright
\def\arraystretch{1.2}
\begin{table*}%
\small
\caption{Generative AI Use Case Examples in Science vs Operations Roles}
\label{table:findings_summary}
\centering
\begin{tabular}{
        p{0.3\linewidth} | 
        p{0.3\linewidth} | 
        p{0.3\linewidth}} 
\hline
\multicolumn{3}{c}{\textbf{Employee Copilot}} \\
\hline
    \textit{Use Case} & \textit{Science Use Case Examples} & \textit{Operations Use Case Examples} \\
\hline
    Writing structured code/text 
    & Writing academic paper introductions, grants, reports, emails, and code  
    & Writing reports, emails, and code \\
\hline
    Extracting insights from large \newline unstructured text data 
    & Extracting insights from scientific \newline literature; querying lab rules and \newline regulations
    & Extracting insights from public data sources such as specified websites; \newline summarizing team data such as surveys; querying lab rules and regulations \\
\hline
\multicolumn{3}{c}{\textbf{Workflow Agent}} \\
\hline
    \textit{Use Case} & \textit{Science Use Case Examples} & \textit{Operations Use Case Examples} \\
\hline
    Initial steps towards automated \newline workflow agents  
    & Operating and extracting data from \newline scientific instruments; automating data analysis pipelines 
    & Automating instrument safety checks \\
\hline 
    Fully automated workflows  
    &  ``AI scientist'' that can generate \newline hypotheses, run simulations, and revise hypotheses based on output with \newline minimal human input 
    & Complex project management, such as prioritizing tasks, creating Gantt charts based on team updates, and long-term planning \\
\hline
\end{tabular}
\end{table*}
\egroup

\subsection{Science and Operations Employee Copilot}
\label{findings:copilot}

We categorize the first set of generative AI use cases described by participants in both Science and Operations roles as \textit{copilot}-style interactions. This means they entail conversational interactions between the user and the AI where the user gets real-time responses to questions posed to the AI. We use the conceptual framing of a copilot in order to arrange our findings around features and affordances this copilot would need to be most useful in a science organization. We note that participants themselves rarely used the term copilot, rather we impose it for conceptual organization of the findings.

At the time of data collection, participants said they were most often using LLMs such as ChatGPT to get help writing structured text that they can easily verify is correct, such as emails and reports. Participants envisioned goal, however, was to use a LLM to extract insights from unstructured text data such as scientific literature or survey results. We group Science and Operations employee responses together in this section since we did not find a large difference in use for copilot-style interactions.

\subsubsection{Current Uses: Writing Structured Code/Text}

Survey and interview participants described numerous examples of structured writing that could be aided by LLMs such as creating emails, reports, grants, and the introduction to academic papers. By structured, we mean types of writing that follow standardized formats. Participants described that they typically already had the content needed for the writing, but they used LLMs to craft the the appropriate tone and format. In the survey, over 50\% of respondents said they had tried using LLMs for each of the following related tasks: \textit{editing human-written text}, \textit{writing code}, \textit{formatting writing}, and \textit{email writing} (Figure \ref{fig:frequencytasks}). Moreover, 20\% of short answer responses included use cases related to writing structured text or code (Table \ref{table:codebook_survey}). 

When writing emails, participants most commonly described using LLMs to help with tone given a specific audience. P3, a senior IT employee, summarized the feelings of many participants saying that he would provide the LLM a document or paragraph and ask it to, \textit{``make this more formal, make this less formal, make this friendlier''} to reduce his own emotional labor. Many participants described wanting to sound ``professional.'' For example, P18, an operations manager, said she asks the LLM to \textit{``make this sound better or make this more professional or make this more succinct.''} Similarly, P21, a scientist manager, said he used ChatGPT to make emails \textit{``more formal and less scientific.''} P16, a scientist, even explained that he could get frustrated but did not want that to come across in his emails so, \textit{``AI writes the email so that I don't have to respond with the emotional outrage I might have.'' } P12, a scientist, noted that as a non-native English speaker, she found this ability to set the tone with the LLM particularly helpful and several survey respondents mentioned using LLMs for language translation.

In addition to writing emails, participants also described how LLMs were helping them write other kinds of structured text such as internal reports. The national lab, like many workplaces, requires specific types of reporting from employees. P17, a scientist manager, described this saying, he was required to write a specific project document and that the national lab had provided \textit{``a guideline that is like a template''} and he felt that \textit{``the AI can prompt you and make sure that you're giving it all of the paragraphs that it needs and then have it summarize the report for you.''} P2, a senior employee in Operations, also found LLMs to be useful in writing internal reports saying, \textit{``Who wants to write these things, right? Who has the time? But... if you take an incident debriefing and ask the system to write a [report] based on that... it did a really nice job.''} 

Scientists also mentioned grants and research paper introductions as other forms of structured writing LLMs could help format. P11, a scientist, described the effort needed to write grant proposals as ``enormous'' and that \textit{``anything that saves time in that process is potentially hugely valuable because it lets the researcher then focus that time on the actual research... I think there's a potentially big advantage if we could get to the point where there's a useful place for LLMs in that process.''} P16, P19, and P22 all thought LLMs could take a research paper draft and summarize the content to draft the introduction to the paper. To reiterate, these were not cases of the LLM generating the research content, but reframing existing content to fit the typical structure of an research paper introduction (in Section \ref{findings:risks_publishing}, we further explore ethical concerns related to AI authorship).

\subsubsection{Envisioned Uses: Extracting Insights from Large Unstructured Text Data}

As science-focused knowledge workers, national lab employees must process significant quantities of unstructured textual data, such as scientific literature or organization regulations. At the time of the study, employees were hesitant to trust generative AI with extracting insights due to fears of hallucinations and reliability, which we return to in the Section \ref{findings:risks_reliability}. If these issues were resolved, however, we found a strong desire among participants to be able to get help from generative AI with managing, organizing, and learning from designated data sources. In the survey, 60\% of respondents said they had at least tried \textit{summarizing literature} (Figure \ref{fig:frequencytasks}) and 20\% of survey free responses mentioned use cases related to querying unstructured text data (Table \ref{table:codebook_survey}).

Multiple interview participants envisioned interacting with unstructured data in a conversational manner. In a representative example, P3, a senior Operations employee responsible for anticipating the software needs of scientists, said \textit{``I think the game-changing aspects would be in interactive conversational simulation modeling.''} He continued with an example where a scientist could \textit{``have brainstorming sessions with the system itself, almost replicating the type of thing that happens in colloquia or in focus discussion groups, but making one of those partners an LLM system.''} In this section, we cover the types of text data participants were most interested in.

\textbf{Scientific Literature:} Scientists, in particular, wanted a tool to help search and summarize scientific literature. P11, a scientist, made a representative comment: \textit{``And like every other academic in the world, we have to do a lot of lit reviews, and that's hugely time intensive and it's just tedious''} and he envisioned that once LLMs became more accurate they could \textit{``get each paper down to a set of bullet points, that could be a huge time saving.''} P22, also a scientist, echoed this perspective and pointed to the increasing amount of scientific literature saying, \textit{``Because nowadays, literature, there's too much literature, too many papers. So that could be very helpful. Maybe it can go out, go to other sources, and then in a particular field, summarize, this week, what happened in this field?''}

\textbf{Public Data Sources:} Lab employees who were not directly involved in writing research papers were also interested in the ability to mine public data sources to extract insights. For example, P4, an Operations employee, wanted to be able to track broader trends online to help with security-related tasks. He described his ideal tool saying that he wanted to be able to create \textit{``a customized library where we can add resources to that library and then basically ping those things [with a LLM].''} P4 described that currently, this kind of public knowledge synthesis for security purposes is done mostly manually. In addition to being able to query data, P7, another security specialist, said he wanted to know how these kinds of public trends related to research being done at the national lab.

\textbf{Team Data:} At the team level, participants wanted the ability to extract key points and notes from meeting transcripts. P10, an IT employee, said she helped process employee requests for approval to use third party tools and that many of these requests were tools that, \textit{``would transcribe the text of the meeting, but it would also create brief notes or overview outlines of what meetings were about and takeaways from that.'' }P9, a lab safety manager, echoed this desire and explained that he was part of a weekly supervisors meeting and he wanted to be able to store and query the meeting transcript for summaries, context on a topic discussed, action items, and other similar questions. Some teams had surveys that they wanted to analyze, for example, P8, an Operations manager, described a situation where her team collected 313 responses from the lab about a safety incident and they used Argo to help identify themes. %

\textbf{Organization Data:} At the organizational level, many participants were eager to be able to more easily search lab-wide rules and regulations, including everything from vacation policies to science lab safety standards. P19, a scientist, had a representative perspective saying, ``\textit{We have a lot of procedures and documents and policies and all that stuff that's spread all over the place. I'd love to see [LLMs] be used for that.}'' Survey respondent S32 wrote that the types of national lab regulations they wanted to search included: \textit{``policies, directives, executive orders, contract requirements.''}  Participants also described the need for combining existing lab datasets. P8, an Operations manager, said, \textit{``right now our systems are very siloed and are not integrated very well... So I see... potentially doing some kind of [LLM] analysis there.''} S37 also pointed to the capability for LLMs not just to search for lab-wide information, but to surface correlations across documents.

\subsection{Science and Operations Workflow Agents}
\label{findings:agent}

We categorize the second set of generative AI use cases described by participants as \textit{workflow agents}. As opposed to a copilot, an agent navigates a complex task autonomously or semi-autonomously and returns the output to the user. In the context of a science organization, we found agents were emerging as a way of driving workflows in both Science and Operations. Scientific workflows included steps such as downloading data from an instrument or database, running multiple data analysis steps, and producing graphs or other visualizations. Operations workflows included tracking if work is progressing on-time, managing procurement processes, and automating common database interactions. 

Tasks related to workflow agents that survey respondents had tried included: \textit{analyzing data} (43\%); \textit{merging, cleaning, formatting data} (35\%); \textit{figure, graph generation} (27\%); \textit{creating synthetic data} (21\%); and \textit{labeling data} (21\%). In the survey free responses, 27\% of answers (equally split between Science and Operations respondents) mentioned workflow automation as a use case for generative AI. 

\subsubsection{Current Uses: Initial Steps Towards Workflow Agents in Science and Operations}

Participants in both Science and Operations described cases where they were testing using LLMs to automate some of their workflow. Participants reported LLMs are already able to automate some of these workflows in a scientific environment to make them more efficient, but generative AI workflow agents are in early stages of development and use. 

\textbf{Science workflows:} Multiple participants described how they were beginning to use LLMs to automate their customized scientific workflows. P21, a scientist, described how he is starting to use LLMs to automate getting results from a specialized instrument. His research uses lensless imaging, in other words, rather than take a picture with optics \textit{``you can do lensless imaging where you... just record how is the light or the x-ray scattered from the sample.''} In order for the data to be useful, P21 explained that researchers use an algorithm to reconstruct the data to create an image, but this is a complex and compute-intensive process that requires \textit{``experts that really massage the data and tune parameters in order to make this work well.'' }He began experimenting with using LLMs to help with this reconstruction process. He used the analogy of editing a photo using Photoshop saying, \textit{``Think about Adobe Illustrator or Photoshop, right? Let's assume you've got a picture of someone and the background isn't great... So yes, you can play with Photoshop until you get that image about right, but with these large language models... where you can interact with the chatbot and say well can you sharpen the image or change the tone or whatever it is that you want.''} While the technology is not fully functional yet, he sees a future where researchers can similarly ask for a scientific ``image'' to be sharpened without manually fine-tuning every parameter.

P14 is a data scientist on the same Science team as P21 where the research used lensless imaging; he explained that he helped researchers use computational imaging techniques and that he was building an automated data analysis workflow using LLMs. He said that part of his data analysis process involved setting over 20 parameters in a script, and even though he wrote the script he still forgot the names of some of the variables, so he set up an LLM to \textit{``convert natural language [parameter names] to Python code or MetaLab code, which directly feeds to the computer for actual data analysis.'' } P14 continued that in addition to generating code, he provided the LLM with knowledge of the analysis techniques he uses, since the LLM's understanding of the techniques on its own was not detailed enough. He said that the knowledge is \textit{``essentially a bunch of text files... a summary of my past experience with different techniques.... whenever I have new data coming in, I want to use the workflow for the new data. And as I process the new data, I pay more attention on my thought process. So in the end I summarize them into the knowledge file.''}  P14 and a colleague were the main users of his automated workflow, however, he said it is \textit{``actually more useful to inexperienced users or people who just started to learn the technique.''} While P14 was optimistic about the ability for LLMs to automate large swaths of his workflow, he said they were still limited to ``narrow environments.'' 

\textbf{Operations workflows:} At this point in time, only the more technically-oriented Operations employees had actually tested creating agent-driven workflows. P1, a safety expert, said he was automating workflows for his group, including building \textit{``an application for performing instrument checks in the morning.'' } He described that since \textit{``a lot of the [instrument checking] software involves these really old scripting-based inputs from the 1960s, no one has a helpful GUI anymore.''} So, he designed a workflow agent where he could \textit{``select, `I want this [chemical element], I want these parameters,' and then just click run''} and the LLM automatically generated a Python script that controlled the old software programs. P1 explained that he did not have any formal training in software engineering but in addition to using a LLM to drive the automation, he also used it to design the Python script: \textit{``I would never have been able to [write the code] without significant time investment, and the fact that I could produce a working app in a couple of days was impressive to me.'' } In addition, P1 described a database his team maintained of relevant chemical elements. He explained how LLMs could be useful for automating searching this database as part of workflows so that his team did not need to export to Excel and do their calculations manually. He wanted to be able to ask the database using an LLM, \textit{``Hey, there was an incident or a fire in this lab. What's the total amount of activity that was in this lab?''} While P1's use cases are specific to his group, the broader ideas he described apply to numerous Operations roles such as: automating old software tools, writing an automation software program without coding skills, and simplifying database searches.

\subsubsection{Envisioned Uses: More Fully Automated Workflow Agents}

Participants in Science and Operations envisioned uses for generative AI that were more fully automated extensions of their current uses. 

\textbf{Science workflows:} Researchers at the national lab have been working towards science-specific large language models that could be integrated into an ``AI scientist'' tool.  As P3 and P20 described, an AI scientist is an advanced workflow agent able to produce scientific simulations that generate new hypotheses that get tested immediately in a feedback loop with minimal human intervention. P20, a software engineer on a Science team, explained his understanding of the initiative: \textit{``[there is] a humongous search space in different science domains, whether it be material science or biology or something like that, and trying to give guidance to researchers so that they don't waste time or it's not just like throwing darts at a dart board.''} 

\textbf{Operations workflows:} Multiple participants in Operations roles envisioned how LLMs could more fully automate their workflows. S60 provided a representative comment that they wanted LLMs to, \textit{``automate more things---auto organize email, create auto-responses to emails, help organize my task list(s) by prioritizing and setting dependencies (can't start task B until task A is complete).''} P17 is a senior manager in a Science division in charge of upgrading a major experiment, however, his day-to-day work aligned more with operations than research tasks.\footnote{We note that in an organization like the national lab, Science and Operations roles can be interrelated, which is part of why studying both sides is important.} During the multi-year project to upgrade the experiment, he said there were safety talks every day that \textit{``would outline which teams were in which areas on which days.''} He envisioned how a workflow agent could use the slides and content from these talks to \textit{``summarize the tasks for all of the teams and create a Gantt chart''} that he could then compare against the timeline that was originally planned in order to \textit{``tell me what took longer and tell me what was faster than what was planned.''} Overall, participants envisioned that workflow agents could automate operations tasks related to communication and project management. 

\subsection{Risks and Concerns}
\label{findings:risks}

In this section we cover the most common risks, concerns, and barriers to using generative AI that participants mentioned. We focus specifically on these topics in the context of participants' work in a science organization. Some interview participants and 14\% of survey respondents mentioned broader social concerns---such unauthorized web scraping for training data and bias in datasets---but since they did not tie these back to implications for their own use of generative AI in their work we do not include them in the discussion.

\subsubsection{Reliability} 
\label{findings:risks_reliability}
Our findings show that the most significant barrier to adoption in a science organization is generative AI's lack of reliability and tendency to hallucinate, as well as the fact that it does not cite source material. Many interview participants and 44\% of survey respondents mentioned reliability concerns. One representative survey respondent (S19) said that LLMs were, ``\textit{not `smart' enough to properly search, cite, summarize literature. Issues include incorrect information and making up fake citations. }'' P4, a lab safety employee, summarized this viewpoint saying, \textit{``I need to be able to tie [an LLM output] back to an authoritative source.''} Citing sources was particularly important for scientists, who noted that current LLMs were not reliable enough for technical information unless they were used by an expert who could spot the falsehoods.

\subsubsection{Overreliance} 
\label{findings:risks_overreliance}
Many interview participants and 21\% of survey respondents were concerned that the introduction of LLM tools would lead to overreliance or inappropriate use cases. Representative survey responses included, \textit{``I'm concerned that people who don't know enough to realize the LLM isn't always correct will use the information as if it's true.'' } (S39) and ``\textit{[LLMs are] a tool and over-confidence in their inherent ability will introduce a bias in the that LLM is `right' or `correct.'}'' (S37). Multiple participants said they were particularly concerned about the use of generative AI in the hiring process, with S29 giving the example, ``\textit{for instance feeding resumes or communications to the LLMs and having them making a [hiring] determination on a more qualified candidate.}'' Participants who felt they understood how generative AI works were not concerned about overreliance issues for themselves, but for others who they thought might not understand. This was the case for both Science and Operations participants. For example, S18 said, ``\textit{I've seen very smart researchers try to use it as a search engine to verify information,}'' which is problematic because LLMs can hallucinate false information.

\subsubsection{Privacy and security for unpublished, classified, and proprietary data}
\label{findings:risks_security}
All of our interview participants and 42\% of survey respondents had some degree of privacy and security concerns with LLMs in their work. P12, a scientist, broadly summed up this perspective saying, \textit{``I do not have fear for the technology itself, just to not know who will use it and who will have access to the data. That seems what I am worried about, especially about the datasets.''} While participants agreed there were concerns about feeding data into commercial LLMs, they differed on exactly where to draw the line, particularly with unpublished academic research.

Lab employees handle both classified data as well as sensitive data such as personally identifiable information (PII), and participants did not feel it was safe to put these kinds of data into commercial LLMs. For example, P13, a scientist, explained \textit{``Parts of the project that I work on are controlled information, so obviously I couldn't be posting anything like that there.'' } P11, a scientist manager, felt the same way about public and commercial LLMs, but thought the national lab's private LLM instance \textit{``might even be okay for most types of controlled and classified information.''} However, he noted more work would be needed to ensure the safety of classified information, \textit{``Of course, there are other use cases where it's more sensitive than that. That would be a whole different conversation, because I have to be running in a separate [secured digital] environment, things like that if we're actually dealing with classified information.''}

Participants disagreed on the extent to which their own unpublished research should be shared with commercial LLMs. At one end of the spectrum, P13, a scientist, did not want to share anything back to companies like OpenAI, saying, \textit{``I would never post, for example, portions of my code and then ask why does this not work? Or something like that.''} In contrast, many participants did feel comfortable sharing code snippets and in fact this was a common use case. The national lab has policies regarding ``export control,'' or controlling how research is published to the public. Several participants brought up that putting an unpublished paper into a commercial LLM might constitute publishing it publicly and thus fall under export control policies, but overall there was uncertainty about this issue. On the other hand, some scientists said that without context their results were useless and that they were planning to publish them soon anyway so it was not an issue. For example, P16, a scientist, said he was not concerned about putting his unpublished research into an LLM because \textit{``if it's proprietary, I don't use [commercial LLMs]. And if it's not proprietary, I'm about to publish it so that everybody can read it.''}

For many participants, using Argo eased privacy and security concerns because it used a private instance of OpenAI's models. For example, P5, a facilities engineer, voiced a common feeling that the lab could be trusted more than companies: \textit{``In a lot of ways I kind of trust [Argo] more than Claude or... ChatGPT. I trust [it] to keep my personal information more secure than I trust a private company.''} Similarly, P6, who works in IT, said he did not have security concerns with Argo because, \textit{``It's contained within our firewall, so I don't have any concerns with that. I think if we would use another LLM that I would have privacy concerns, but not with [our instance] in particular.''} S19 said, \textit{``My work involves sensitive information and novel experiments, and so I am apprehensive to provide too much detail to [non-Argonne National Lab] LLMs.''} At this time, the private instance also does not store chat history. P7, a lab safety employee, said this was good for privacy and security, explaining, \textit{``I think the best thing is that [chat history] doesn't get stored anywhere because... even in [the national lab] there's going to be probably [several thousand] people who could access it. So with every different use case in science, admin, and stuff like that. So I think making it so if someone puts something in that maybe shouldn't be, the fact that it won't be there is good from a more privacy focused [angle].''}

However, this strategy of promoting Argo could backfire if employees prefer other more advanced LLM models that are available. At the time of the study, OpenAI had released GPT-4 but only GPT-3.5 was available through the national lab's system. P9, a lab safety manager, explained this saying, \textit{``I think the capability of that if we're hosting it here on site... is just so much more limited to what's there. The natural attractiveness of going outside of [the national lab's] intranet is just obvious based on capability.''} P21, a scientist manager, echoed this saying he preferred using ChatGPT: \textit{``I've been using chatGPT as opposed to the [Argo] thing. So far, I have not used it with any data where we have any concerns with regards to security, privacy, export control, or any of these topics.''} While participants who were using public models said they were being careful, from an organizational perspective this could be a security threat. 

\subsubsection{Academic Publishing in the Era of LLMs}
\label{findings:risks_publishing}
Many scientists and 20\% of survey respondents were concerned about researchers using AI to write academic papers. However, they did not agree on exactly where to draw the line on acceptable vs unacceptable use. On the one hand, everyone agreed that generating fake research was unacceptable. For instance, P21, a scientist manager, said, \textit{``Some individuals... just create fraud... AI and machine learning opens that up to a whole new level because you may be able to really create papers at a push of a button.''} Similarly, P19, a scientist, said, \textit{``I think there is also an ethical concern of people using something like a large language model to write a paper and then publishing it and claiming it as their own work. So there's a plagiarism aspect there.''} Another scientist, P16, said he had been asked to review some AI-written work for a journal and had suggested it be rejected. 

Part of the issue, participants suggested, is that if something in the paper turns out to be incorrect it is not clear who to blame. Also, if the LLM suggests something useful, it is not clear how to cite it. For example, P19, a scientist, said, \textit{``If I write a paper and I submit it and it turns out that I said some stuff that was wrong, well at least there's a person we can have that conversation with. If it comes from a chatbot, then what do you do with it?''}

Most participants felt some level of paper editing with a LLM was acceptable, but the line was fuzzy. P13, a scientist, outlined this tension felt by many saying he was ``troubled'' by \textit{``the idea of generating content that you pass off as your own''} while at the same time he felt that if ChatGPT was just used to help minimally refine a researcher's original idea then it was overkill to \textit{``put a disclaimer at the front of your work that says this was partially or fully AI assisted.''} Overall, participants saw a role for LLMs in writing research papers, but we found that it was still an open question as to exactly how much assistance would be viewed as acceptable. 

\subsubsection{Impact of Generative AI on Jobs} 
\label{findings:risks_jobs}
Participant opinions were mixed on the impact generative AI would have on jobs. Many interview participants brought up the topic, and we asked managers about it directly. While it was only mentioned by 5\% of survey respondents, some felt strongly it was important. For the most part, participants saw generative AI as a tool that would be helpful to but not replace scientists, but there was some concern for jobs that required less scientific expertise, such as Communications or IT roles. Many managers believed the skill sets they hired for would change but not the total number of workers.

Participants in technical or specialized roles in Science and Operations tended to view generative AI as a tool that could speed up their infinite to-do lists, but that would not replace them. For example, P17, a scientist manager, said \textit{``[generative AI] will just be another tool like FEA [finite element method] analysis, right? FEA analysis is faster than doing it by hand and more thorough [and LLMs will] be like that.''} Another scientist manager, P21, similarly said, \textit{``It's a new technology that will have a very significant impact on the workforce, but essentially it's like a tool.''} Likewise, P16, a scientist, said \textit{``It's giving me new tools to use.''} Most technical workers were not concerned about losing their jobs due to generative AI. P9, a safety manager with a technical background, summarized this saying the efficiency gained from using LLMs, \textit{``opens up time to do other things that wouldn't have gotten done''} and that he was \textit{``completely comfortable with the idea that it's better at [some tasks]... I have no concerns of that taking my job away.''} P16, a scientist, said that having a LLM be able to do programming tasks was like \textit{``having a free junior scientist''} that his group would not have had the money to hire a human for in the first place. However, he said much of his job required being on-site at a scientific experiment \textit{``to make sure that a laser hits the right spot''} and he felt \textit{``AI is never going to be able to do that well, at least not for a very long time.''} P18, who works in Operations, said, \textit{``Nothing [in our work] is general. We're not doing general, we're solving a problem based on a specific set of concerns''} and therefore that she was not concerned \textit{``about it taking over people's creativity of their work application, yet at least.''} Only one technical participant, P14, thought AI might take his job as a data scientist one day but seemed thrilled by the possibility rather than afraid saying, \textit{``My dream is to develop a fully automated workflow that kind of replaces me... Honestly, I think a lot of my job can be replaced by LLM given enough time and training.''}

On the other hand, numerous participants in both Science and Operations were more concerned that AI might take over less technical or specialized jobs. For example, P6 works in IT and his job was largely managing existing automation and fixing issues as they arise: \textit{``I will update correct data in the database tables, troubleshoot issues with some applications for production support, upgrade small commercial off-the-shelf products.''} Given that it is conceivable LLMs will be able to automate some of these tasks, he was concerned, saying \textit{``I think that somewhere in the near distant future they're going to put people out of work... you're going to have somebody that is in my [older] age group that's maybe stubborn and they don't want to learn technology.''} P16, a scientist, thought AI might take jobs that are largely reading or writing based saying if \textit{``your job is to attend meetings and write emails, AI could replace 90\% of what you do.'' } P19 was concerned about AI taking jobs but not research jobs that required ``technical writing.'' S18, who works on the Operations side in a Communications role, said she was alarmed to see a recent LLM-generated publication that she felt plagiarised her ideas without credit. She also felt her work was devalued by researchers who did not understand the effort required for science communication writing, saying LLMs led \textit{``to a lack of respect for creative arts or content-generating colleagues.''} She also warned that \textit{``it's only a matter of time before research-based content experiences the same [impact from generative AI-authored work.]''} 

When we asked managers how they thought generative AI might change hiring, for the most part they said it would change the kinds of skills they looked for in workers but not the total number of workers. Technical managers such as P21 made comments such as, \textit{``I expect that more and more people that we hire will have literacy in AI and machine learning.''} P21 continued saying, \textit{``we typically have scientists that run our instruments and collect, acquire and reconstruct data and help interpret it. I suspect we may need a few less of those, but not dramatically. So instead we will need more sort of AI-based people that are able to tweak, let's say large language models.''} Non-technical managers were also adjusting the skill sets they looked for. P18, an Operations manager, said \textit{``I would've probably told you 10 years ago written communication was a better skill. Now I'm looking for people who can actually have a conversation on the phone... So I do think that skill sets are going to continue to change.''} Overall, there was more concern for non-technical than technical roles, but managers felt that the skills they valued in all roles would be modified based on generative AI's capabilities.

\section{Discussion}

Overall, our findings suggest that participants were familiar with generative AI technology and experimenting with it for a range of tasks from writing reports to automating data science pipelines. Argo usage showed that the number of early adopters for generative AI at the national lab has been growing for both Science and Operations users. These trends, when paired with our interview data, suggest that generative AI tools can add value to science organizations. However, these tools will be the most useful if they are designed to solve real, domain-specific problems, and hurdles such as reliability and accuracy must be improved. In addition, there are risks particular to science organizations that must be addressed. In this section, we discuss design and organizational policy recommendations and ideas for future research directions.

\subsection{Organization Copilots and Workflow Agents: Design Recommendations and Future Directions}

Given the significant overlap between what Science and Operations employees reported looking for in a copilot, we argue a single organizational copilot is needed. However, different domains have unique needs and we see a future where specialized copilots can be designed for different types of workplaces. Based on participant responses, such a \textbf{copilot could be integrated into organization email services} to help workers with tone, with particular support for people for whom English not their first language. In addition, a copilot could have \textbf{knowledge of common report and grant structures and include pre-programmed prompts to help workers put these together} more quickly. It is also \textbf{critical for a copilot to have access to internal organization data} and \textbf{accurately pull information from academic literature and technical documentation with citations}, allowing for interactions such as scientific brainstorming and automatically surfacing correlated data. Moreover, synergy between ideas can advance science more quickly and a copilot could \textbf{help identify common interests between Science and Operations teams}. Lastly, there could be a way for teams to \textbf{upload team-specific frequently asked questions (FAQs) to the copilot} to reduce email burden from common questions. While some of these affordances might be useful in many organizations, science organizations have particular technical context that a copilot must understand for it to be useful.

While organizations could use a single copilot, our findings showed workflows tended to be highly customized depending on the team and project. Organizations that want to support their employees in creating customized workflow agents could \textbf{provide scaffolding and templates for helping workers with best practices for creating an agent}. Organizations could also provide employees with strategies for \textbf{giving generative AI agents scientific or technical context}. Organizations might want to consider \textbf{a way for workers to share agents they have made with others in the organization}. One type of agent we see as being important in science organizations are \textbf{agents for interacting with and operating complex scientific instruments}, an area that merits more research on technical and design considerations for this task. Another type of agent that could be tweaked for both Science and Operations employees are \textbf{agents for the continuous monitoring and summarization of designated information sources} including academic literature and online sources.

Future work could focus on honing in on industry-specific copilot designs, building on prior literature on specific tasks such as writing and programming \cite{Lee_etal_2024, Russo_2024}. For example, answers are needed for what interaction timing is best for organization copilots, how employees can find out about their capabilities, and how complex data sources like science literature can be organized and queried. In addition, future work could look at how to craft a generalised basic workflow agent or template agent that can be adapted by employees in multiple roles and with a range of technical skills. More work is also needed on creating operations workflow agents for common processes around email organization and project management. This builds on literature on AI agents that extends prior to generative AI's development, along with more recent research on generative AI workflow agents in science and elsewhere such as \cite{Dearing_Prince_2023, Drosos_etal_2024, Suh_etal_2023}. Lastly, future work can also measure the extent to which generative AI is being adopted in organizations as the technology becomes more developed.

\subsection{An Organizational Approach to Generative AI Risks}

Based on our findings around organizational risks and concerns with deploying generative AI to employees, we outline directions to mitigate these issues in a science organization. 

\subsubsection{Policies for Publications and Citations} Our findings show that employees have a range of opinions on publication policies and citations when generative AI is used and many were confused about how to proceed. Science organizations, as well as organizations that publish writing to the public, should create clear policies for employees (accounting for all types of roles) around appropriate use of generative AI in writing and citation practices and communicate these policies effectively. While there is no ``right answer,'' organizations can help guide staff so that everyone is working with a shared understanding of what is appropriate use as well as how to cite AI-generated content. Given that we also found significant concerns around reliability and overreliance on generative AI tools (such as hallucinations, lack of scientific or technical knowledge, and overconfidence in incorrect answers), organizations should also prepare for cases where AI-written content includes a falsehood. This may be particularly problematic for science organizations that rely on public trust \cite{Morris_2023} and grant funding. HCI researchers can help guide what these policies should be and also work on designing technical solutions such as studying how watermarks \cite{Regazzoni_watermarking_2021} for generative AI content could work in practice in an organization.

\subsubsection{Mitigating Privacy and Security Risks} A key organizational threat, especially for organizations like a national lab that deal with confidential and even classified information, is managing privacy and security with generative AI, a topic that has little research thus far. At this time, many employees are experimenting with both commercial LLMs such as ChatGPT and lab-specific Argo. Participants told us that they are careful and many described redacting sensitive information before placing it in a LLM, however, this means different employees have different ideas about what is acceptable data to share. Science organizations, in particular, need policies around sharing unpublished academic research, research code, and organization-specific information. We also found many participants felt more comfortable, from a privacy and security perspective, using Argo since no queries were stored or shared. This suggests that a viable option is for organizations to make these kinds of internal generative AI systems available for employees in order to mitigate the need to go to outside products. Organizations must, however, make sure their internal options are competitive with external ones since otherwise employees may continue to use external generative AI with better features. Even with an internal LLM that has access to organization data, given employees have different levels of information access, future work could investigate how to design this into the system, particularly for classified information.

\subsubsection{Transparency Around Future Hiring and Skills Needed} It is important for organizations to understand that some employees may be concerned about the future of their jobs given the introduction of generative AI. Echoing earlier findings \cite{Morris_2023}, at this time, scientists largely felt their jobs were safe. However, the ``AI scientist'' project at the lab as well as managers' comments that skills they were hiring for would shift due to generative AI indicate that scientific roles may also be impacted. From an organizational standpoint, we recommend leaders make it clear to employees how they see AI impacting future hiring and what skills will be valued in both operations and knowledge-specific roles. 

\subsection{Limitations}

This paper provides a case study of a single science organization. Given how little is currently known about organizational use of generative AI assistants, findings from this case study are applicable to a broad range of knowledge work institutions such as those with knowledge specialists (e.g. scientists, lawyers, etc.) and operations workers; in addition, future research should study a variety of science and other knowledge work organizations. When studying usage in the organization, we did not have access to the number of employees accessing commercial LLMs and so our usage data for Argo under-counts total LLM usage. Our participants skewed toward white men, and while this is reflective of the organization a greater diversity of participants might have offered new insights, and future work could correlate specific use cases and risks with demographic features. Our participants also tended to be early adopters of generative AI who were interested in the subject and may not reflect the opinions of employees with little background or interest in the technology.

\section{Conclusion}

In this paper, we study the practical, real-world applications and perceived risks of generative AI use across Science and Operations teams in a multidisciplinary science and engineering research center, Argonne National Lab. To understand current and envisioned generative AI use cases and privacy, security, ethics, and other concerns surrounding generative AI in a science organization, we report on usage statistics for the first release of a private instance of GPT-3.5 called Argo at the lab, a survey (\(N=66\)), and in-depth interviews (\(N=22\)). We find that there is an upward trend of Argo users, split between both Science and Operations employees, although use is largely experimental at this time. Uses cases fall into either a \textit{copilot} or \textit{workflow agent} generative AI modality. Risks include reliability, overreliance, privacy and security, the impact on academic publishing, and concerns around generative AI taking jobs. We end with recommendations for organizations interested in implementing generative AI systems and for HCI researchers working on crafting these systems.

\begin{acks}
We thank the anonymous reviewers for their valuable feedback on this paper, as well as the professionals who shared their time and experiences with us. This project was funded by the University of Chicago Data Science Institute's AI+Science Research Initiative.
\end{acks}

\bibliographystyle{ACM-Reference-Format}
%%% -*-BibTeX-*-
%%% Do NOT edit. File created by BibTeX with style
%%% ACM-Reference-Format-Journals [18-Jan-2012].

\appendix
\newpage
\section{Appendix}

\subsection{Survey Instrument}
\label{appendix:survey}

\begin{enumerate}
    \item How familiar are you with large language models (LLMs) such as ChatGPT, Argo, etc.? [Very unfamiliar, Unfamiliar, Somewhat familiar, Familiar, Very familiar]
    \item Which LLMs have you used before? [Argo, ChatGPT, Claude, Gemini, LLaMA, Other]
    \item How often do you use LLMs as part of your work? [Never, Rarely, Sometimes, Often, Always]
    \item How often do you use LLMs for personal use? [Never, Rarely, Sometimes, Often, Always]
    \item Where do you get LLM prompts? [I don't use LLMs, I write my own prompts, I use pre-programmed prompts developed by Argonne National Lab, Other]
    \item To what extent would you agree/disagree with the following statement: LLMs have become an essential part of my workflow. [Strongly disagree, Disagree, Neither agree or disagree, Agree, Strongly agree]
    \item How often do you use LLMs for the following Argonne National Lab-related work tasks? (Some tasks may overlap, which is OK.) [Never, Rarely, Sometimes, Often, Always]
    \begin{enumerate}
        \item Labeling data
        \item Merging / cleaning / formatting data
        \item Analyzing data
        \item Creating synthetic data
        \item Summarizing literature
        \item Editing human-written text
        \item Formatting writing
        \item Figure / graph generation
        \item Grant writing
        \item Email writing
        \item Writing job descriptions
        \item Science communication for the public
        \item Writing code
        \item Learning about APIs or programming syntax
        \item Feedback on experimental design
    \end{enumerate}
    \item Can you envision other ways LLMs might help with your work? [Short response]
    \item Please describe issues you have encountered using LLMs in your work, if any: [Short response]
    \item Please describe ethical concerns you have about using LLMs in your work, if any: [Short response]
    \item Please describe privacy/security concerns you have about using LLMs in your work, if any: [Short response]
    \item Demographics
    \begin{enumerate}
        \item What is your Argonne National Lab division code?
        \item What is your role?
        \item How long have you worked at Argonne National Lab?
        \item What is your age?
        \item What is your gender?
        \item What category best describes you? [Race/Ethnicity]
        \item What is your highest level of education?
    \end{enumerate}
    \item Provide email for interview sign-up [Optional]
    \item Is there anything else you'd like to share about using generative AI and large language models in the workplace?
\end{enumerate}

\subsection{Interview Protocol}
\label{appendix:interview_protocol}
\begin{enumerate}
    \item What is your job and what are some typical tasks you do at work?
    \item Have you tried using large language models (LLMs) for any tasks? \textit{Go through each task in detail. Prompt based on survey responses.}
    \begin{enumerate}
        \item Could you walk me through an example of the task?
        \item To what extent have LLMs been helpful?
        \item When using an LLM to do [task] doesn’t work, what goes wrong?
        \item Is there any data involved, if so what?
        \item Do you have a go-to prompt for this task? How did you create the prompt?
        \item Would you benefit from guidance on developing effective prompts for this task?
    \end{enumerate}
    \item Are there any tasks you’d like to use LLMs for but haven’t?
    \item Do you have any concerns about using the LLM to do these tasks?
    \item Do you have any privacy and security concerns regarding LLMs?
    \item Do you have any ethics concerns regarding LLMs?
    \item Do you have any hopes or fears for LLMs more broadly?
    \item Which LLMs do you prefer and why?
    \item Have you come across any surprising or weird results when trying LLMs?
    \item To what extent do you see LLMs impacting hiring on your team? [Managers only]
    \item What are you hearing from other people you work with about generative AI?
    \item The national lab is developing a menu of pre-programmed LLM prompts. What kind of support would be most helpful for your work?
\end{enumerate}

\subsection{Interview Codebook}
\label{appendix:codebook_interview}

See Table \ref{table:codebook_interview}.

\begin{table*}[t]
\small
\caption{Interview Codebook}
\label{table:codebook_interview}
\centering
\begin{tabular}{p{0.35\linewidth} | p{0.6\linewidth}}
\hline
 Code & Code Description\\
\hline
\textbf{Current/envisioned uses for LLMs*} & Current or envisioned use cases for LLMs  \\
 Writing structured text/code & Use cases related to writing structured text/code \\
 Query unstructured data & Use cases related to querying unstructured data sets \\
 Workflow automation & Use cases related to workflow automation \\
 Other & Other use cases \\
\hline
\textbf{Issues or ethics concerns at work} & LLM concerns related to ethics in a science/work context \\
Reliability & Concerns about whether LLMs are reliable, trustworthy, etc.\\
Overreliance & Concerns about people relying too much on LLMs\\
Academic publishing & Concerns about academic and science communication publishing\\
AI taking jobs & Concerns about the impact of LLMs on human jobs\\
Social concerns & Broader social concerns that participants do not tie directly to their work\\
No ethics concerns at work & Participant did not have ethics concerns about LLMs at work\\
\hline
\textbf{Privacy/security concerns at work} & LLM concerns related to privacy in a science/work context \\
Concerned & All concerns related to privacy and security\\
Not concerned if using organization tools & Participant not concerned about privacy/security as long as they were using LLMs officially designated secure by their organization\\
Concerned but prefer commercial LLMs & Participant felt organization-designated LLMs were safer but preferred the features available in commercial LLMs\\
No privacy/security concerns at work & Participant did not have privacy/security concerns about LLMs at work\\
\hline
\textbf{Prompt writing}** & How does the participant write prompts, do they use any resources, do they re-use prompts  \\
\textbf{Weird LLM results}** & Weird or surprising results participants have gotten from interacting with a LLM \\
\textbf{Preference between LLMs}** & Reasons why participant prefers (or not) some LLMs over others. This could be general preference or task specific preference. \\
\hline
\end{tabular}
\begin{tablenotes}
\item *Bolded codes represent our initial codebook. Non-bolded codes are thematic sub-codes that were identified after a second round of coding.
\item **We did not include results from these codes in the paper since they did not align with our research questions.
\end{tablenotes}
\end{table*}

\end{document}